\documentclass[footinbib,fleqn,aps,prb,twocolumn,superscriptaddress,reprint,floatfix]{revtex4-2}
\usepackage{xr}
\usepackage[]{amsmath}
\usepackage{amssymb}
\usepackage{graphicx}
\usepackage[english]{babel}
\usepackage{microtype}
\makeatletter
\let\l@en\l@english
\makeatother
\usepackage{color}
\usepackage{tabularx}
\usepackage{mathtools}
\usepackage{algpseudocode}
\usepackage{enumitem}
\usepackage{multirow}
\usepackage{epstopdf}  
\usepackage{adjustbox}
\usepackage{bm}

\usepackage{makecell}
\usepackage{hyperref}
\usepackage[version=3]{mhchem}
\usepackage{tikz}
\usetikzlibrary{shapes,snakes}
\usetikzlibrary{arrows}
\usepackage{threeparttable}

\newcommand{\methodhead}[1]{%
  \par\medskip\noindent\textbf{#1.}\enspace\ignorespaces
}

\begin{document}

\title{Unveiling the Complex high-pressure phases of Sodium at the Melting Curve Minimum via Data-Drive Crystal Structure Prediction.}
  
\author{Zhenyu Wang}
\thanks{These authors contributed equally to the work: Zhenyu Wang and Chengqian Zhang.}
\affiliation{{Key Laboratory of Material Simulation Methods \& Software of Ministry of Education and State Key Laboratory of High Pressure and Superhard Materials, College of Physics, Jilin University, Changchun 130012, P.R.~China}}

\author{Chengqian Zhang}
\thanks{These authors contributed equally to the work: Zhenyu Wang and Chengqian Zhang.}
\affiliation{DP Technology, Beijing 100080, P.R.~China}
\affiliation{College of Engineering, Peking University, Beijing 100871, P.R.~China}

\author{Xiaoyang Wang}
\thanks{Corresponding Author: Jian Lv, lvjian@jlu.edu.cn; Xiaoyang Wang, wang\textunderscore xiaoyang@iapcm.ac.cn}
\affiliation{Laboratory of Computational Physics, Institute of Applied Physics and Computational Mathematics, Fenghao East Road 2, Beijing 100094, P.R.~China}

\author{Haiyang Niu}
\affiliation{{State Key Laboratory of Solidification Processing, International Center for Materials Discovery, School of Materials Science and Engineering, Northwestern Polytechnical University, Xi'an, P.R.~China}}

\author{Jian Lv}
\thanks{Corresponding Author: Jian Lv, lvjian@jlu.edu.cn; Xiaoyang Wang, wang\textunderscore xiaoyang@iapcm.ac.cn}
\affiliation{{Key Laboratory of Material Simulation Methods \& Software of Ministry of Education and State Key Laboratory of Superhard Materials, College of Physics, Jilin University, Changchun 130012, P.R.~China}}

\author{Han Wang}
\affiliation{Laboratory of Computational Physics,
  Institute of Applied Physics and Computational Mathematics, Fenghao East Road 2, Beijing 100094, P.R.~China}
\affiliation{HEDPS, CAPT, College of Engineering, Peking University, Beijing 100871, P.R.~China}

\date{\today}
	
\begin{abstract}
Sodium exhibits a remarkable concentration of complex crystalline phases near its melting-curve minimum, yet the atomic arrangements of several experimentally observed phases remain unresolved. Here, we combine CALYPSO crystal structure prediction with Deep Potential models developed through concurrent learning to explore this structural landscape. Searches over more than 1.6 million configurations recover the previously reported $I4/m$ structure of $tI50$ and identify candidate atomic arrangements for $oC120$ and $aP90$, together with three additional low-enthalpy structures, $tI52$, $oP72$, and $oP92$. The proposed candidates reproduce key features of the experimentally indexed cell geometries. First-principles enthalpy calculations, Gibbs free-energy comparisons, and solid--liquid coexistence simulations reveal close competition among these structures and established sodium phases. Near the calculated reference melting minimum, $tI50$ is favored over $cI16$ and $oP8$, whereas $oC120$ and $aP90$ remain closely competitive, with their relative stability sensitive to model accuracy. These results provide testable structural models for unresolved experimental phases and reveal multiple closely competing low-enthalpy configurations near sodium's melting minimum
% predictions or searches
\end{abstract}
\maketitle

\section{Introduction}

At ambient conditions, alkali metals are prototypical simple metals, with highly symmetric crystal structures and nearly-free-electron behavior. Compression, however, reshapes their electronic structure and chemical bonding\cite{HighPressureReview2017NRM}, leading to a rich sequence of structural transitions and the emergence of complex crystalline phases\cite{HighPressureElems2006CSR}. This structural diversity is accompanied by unusual physical properties, including superconductivity\cite{LiSC2002Science,LiSC2002Nature,RbSC2019PRB}, metal–semiconductor\cite{Marques_oC40_oC88_oC24PRL2011,LvoC40Li2011PRL,LiBandGapNature2009} and metal–insulator\cite{MaNa2009Nature} transitions, and anomalous melting curves\cite{Mungo_lithium_300K_crystal_prl_2019,Christophe_lithium_oC88_oC24_oC40_190K_natphy_2011,Schaeffer_lithium_300K_disorder_prl_2012}. Such structural and physical complexity stands in striking contrast to their ambient-pressure simplicity. Determining the atomic arrangements of these complex phases is therefore essential to understanding the unusual behavior of alkali metals under compression.

As a notable example, sodium exhibits a succession of pressure-induced structural transitions at room temperature\cite{NaPhaseDiagram2008Science}, evolving from the simple bcc and fcc structures into \(cI16\) at approximately 102 GPa, \(oP8\) at 118 GPa, and the incommensurate host–guest \(tI19\) phase at 125 GPa. Further compression to about 200 GPa produces the optically transparent, insulating \(hP4\) phase, in which valence electrons become localized in the lattice interstices\cite{MaNa2009Nature}. These interstitially localized electrons act as anions, making \(hP4\) sodium an elemental electride.

Sodium also exhibits an anomalous melting curve: its melting temperature initially rises to approximately 1000 K at 31 GPa, then falls steeply to a minimum near 300 K at 118 GPa before increasing again upon further compression\cite{NaPhaseDiagram2008Science}. Around this minimum, single-crystal X-ray diffraction experiments revealed seven crystalline phases within a narrow pressure–temperature region, with small changes in pressure or temperature inducing transitions between distinct structures. Among these seven phases, \(cI16\), \(oP8\), and \(tI19\) were structurally characterized. The remaining four were indexed as large-unit-cell phases with Pearson symbols \(tI50\), \(oC120\), \(aP90\), and \(mP512\), but their detailed atomic arrangements remained unresolved for nealy two decades. A recent computational study proposed an \(I4/m\) structure as a candidate for \(tI50\), marking progress toward resolving these complex phases\cite{YaoNa2025CommChem}. Nevertheless, the diversity of phases observed within this narrow pressure–temperature region motivates a systematic exploration of possible atomic arrangements and their relative stability, to identify plausible structures for the unresolved phases and clarify their competition near sodium's melting minimum.

Such an exploration, however, presents a substantial computational challenge. The large unit cells and low symmetry of these phases greatly expand the configurational space to be searched, making extensive crystal structure prediction based directly on density-functional theory (DFT)\cite{DFT1964PR,DFT1965PR} computationally demanding. Establishing their relative stability near melting introduces a further challenge, as finite-temperature free-energy calculations require extensive sampling beyond structural optimization at zero temperature.

Machine-learning interatomic potentials offer a practical means of addressing both demands by approximating first-principles potential-energy surfaces at substantially reduced computational cost, enabling large-scale structure searches and finite-temperature simulations. This strategy\cite{ConcurrentLearning2024PRB} has demonstrated its effectiveness in dense lithium, where experiments revealed evidence for unidentified crystalline phases near the melting-curve minimum. Machine-learning-assisted structure searches predicted four complex structures containing up to 192 atoms per unit cell that are energetically competitive with known lithium phases, providing plausible candidates for the observed phases\cite{WangLi2023NC}. These advances offer a promising route to resolving the complex high-pressure structures of sodium and clarifying their competition near its melting minimum.

In this work, we investigate sodium's unresolved phases near the melting minimum by combining CALYPSO\cite{calypso2010PRB,calypso2012CPC,Lvcalypso2012JCP,calypso2022JCP,WangHPCSD2026SciBull} crystal structure prediction with Deep Potential (DP)\cite{ZhangDP2018PRL,HanDP2018CommCompu} models developed through concurrent learning. This approach makes extensive exploration of large, low-symmetry cells feasible, allowing us to examine more than 1.6 million configurations. Our searches yield candidate atomic arrangements for the experimentally reported \(tI50\), \(oC120\), and \(aP90\) phases, with lattice geometries consistent with available experimental constraints. We further predict three energetically competitive phases, \(tI52\), \(oP72\), and \(oP92\), extending the structural diversity of sodium in this region. First-principles enthalpy calculations, finite-temperature Gibbs free-energy comparisons, and solid–liquid coexistence simulations probe their stability and melting behavior. Several candidate structures compete with established sodium phases within an energy window of only a few meV per atom, revealing a landscape of closely competing crystalline states near the melting minimum. These findings provide concrete structural models for longstanding experimental observations and show how multiple complex atomic arrangements can compete within a narrow pressure–temperature region.

\section{Methodology}
We combined machine-learning-assisted crystal structure prediction with first-principles calculations and finite-temperature simulations to investigate the complex high-pressure phases of sodium near its melting minimum. Three DP\cite{ZhangDP2018PRL,HanDP2018CommCompu} models were developed for different computational tasks: DP-MP for a preliminary determination of the melting minimum, DP-CSP for crystal structure searches, and DP-FE for finite-temperature thermodynamic calculations. The training datasets were progressively expanded to cover the configurations relevant to these tasks.

\methodhead{DFT calculations}
DFT\cite{DFT1964PR,DFT1965PR} calculations were performed to construct the training and testing datasets, refine the predicted structures, and calculate their relative enthalpies. We employed the Perdew–Burke–Ernzerhof exchange-correlation functional within the generalized gradient approximation\cite{PBEGGA1994PRL,PAW1994PRB} and SG15 optimized norm-conserving Vanderbilt pseudopotentials\cite{ONCV2013PRB} for sodium. Methfessel–Paxton smearing was applied with a width of 0.0146 Ry. A plane-wave energy cutoff of 100 Ry and a Monkhorst–Pack \(k\)-point grid spacing of \(0.07\ \text{\AA} ^{-1}\) were used. These settings ensured convergence of the total energies, atomic forces, and virial tensors to within 1 meV, \(10\ \text{meV/\AA}\), and 10 meV/atom, respectively. Calculations for dataset construction were carried out using ABACUS\cite{ChenABACUS2010JPCM,LiABACUS2016CMS}, whereas structural relaxations and enthalpy calculations for the predicted phases were performed using Quantum ESPRESSO\cite{QE2009JPCM,QE2017JPCM} with the same functional, pseudopotentials, and numerical settings.

\methodhead{Construction of DP models}
The training datasets were constructed using the concurrent learning scheme implemented in the Deep Potential Generator (DP-GEN) platform\cite{DPGEN2020CPC}. Starting from an initial dataset, DP-GEN iteratively performed model training, configuration exploration, and DFT labeling. In each iteration, four DP models were trained using DeePMD-kit\cite{Deepmdkit2018CPC} with the same dataset and hyperparameters but different random seeds. One model was used to explore the configuration space, while the ensemble was used to evaluate the sampled configurations. The maximum standard deviation of the predicted atomic forces among the four models served as the model-deviation indicator. Configurations with model deviations between prescribed lower and upper bounds were selected for DFT calculations, and the resulting data were added to the training dataset for the next iteration. Molecular-dynamics exploration was performed using LAMMPS\cite{LAMMPS2022CPC} interfaced with DeePMD-kit\cite{Deepmdkit2018CPC}, while structure-search exploration used CALYPSO\cite{calypso2010PRB,calypso2012CPC,Lvcalypso2012JCP,calypso2022JCP,WangHPCSD2026SciBull} within the DP-GEN\cite{DPGEN2020CPC} workflow.

For DP-MP, the initial dataset consisted of DFT-relaxed structures of the experimentally established bcc, fcc, \(cI16\), and \(oP8\) phases, together with configurations generated by perturbing their atomic positions and cell vectors. Subsequent DPMD exploration covered pressures of 110–180 GPa and temperatures of 10–1000 K, sampling both crystalline and liquid configurations. The resulting model was used to obtain a preliminary estimate of the pressure region around the melting minimum.

%%%%%%% Insert Figures %%%%%%%%%
\begin{figure*}[htbp]
\begin{center}
\includegraphics[width=0.9\textwidth]{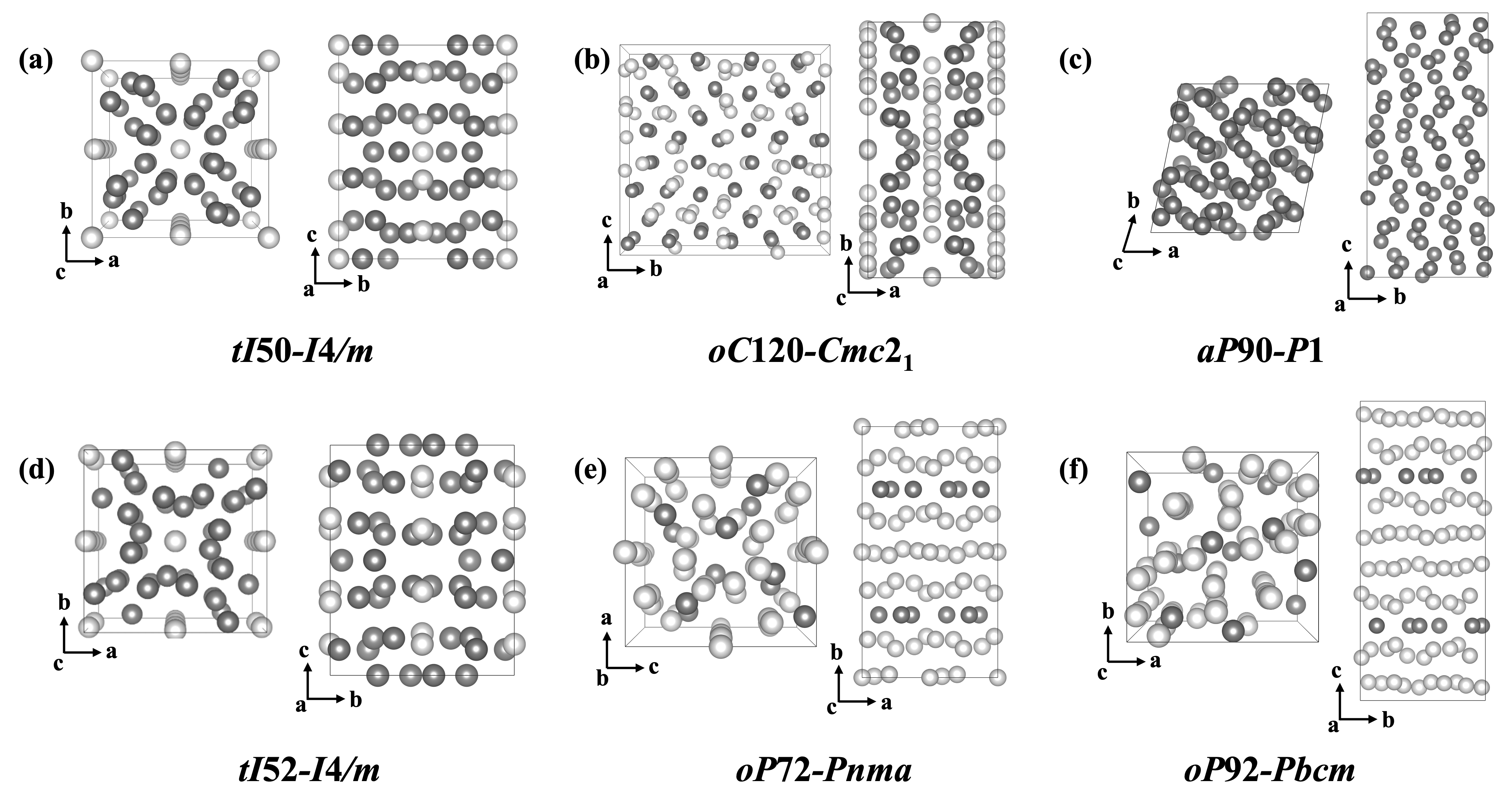}\\[5pt]  % insert figure
\caption{{\textbf{Predicted Structures of Experimentally undetermined phases}} (a) tI50-I4/m. (b) oC120-Cmc21. (c) aP90-P1. (d) tI52-I4/m. (e) oP72-Pnma. (f) oP92-Pbcm.
}
\label{fig:Fig1}
% \vspace{-0.4cm} %调整图片与上文的垂直距离
\end{center}
\end{figure*}
%%%%%%%%%%%%%%%%%%%%%%%%%%%%%%%%%

The DP-CSP dataset was constructed by augmenting the DP-MP dataset through the CALYPSO-based concurrent learning scheme described in our previous work\cite{ConcurrentLearning2024PRB}. Here, configuration exploration was performed through short structure searches, and configurations along the local relaxation trajectories were evaluated using the model-deviation indicator. This procedure extended the training data beyond configurations sampled by molecular dynamics of known phases to include atomic environments encountered during structure searches. The resulting DP-CSP model was then used for extensive searches for complex sodium structures.

For DP-FE, the combined DP-MP and DP-CSP datasets were further supplemented through DPMD exploration initiated from the predicted \(tI50\) structure and commensurate approximations to the \(tI19\) phase. The exploration covered pressures of 130–180 GPa and temperatures of 10–450 K. Additional DFT data for relaxed structures and finite-temperature configurations of \(cI16\), \(oP8\), \(tI19\), \(tI50\), \(oC120\), and \(aP90\) were subsequently incorporated to refine the model for thermodynamic calculations.

DP-CSP used the smooth two-body embedding descriptor, whereas DP-MP and DP-FE employed a hybrid of two-body and three-body embedding descriptors\cite{Se}, with cutoff radii of 6 and 3 \AA, respectively. The two-body embedding, three-body embedding, and fitting networks contained three hidden layers with sizes of (25, 50, 100), (4, 8, 16), and (240, 240, 240), respectively. Further details of dataset construction, model-deviation thresholds, and training settings are provided in Supplemental Sec. I.

\begin{table*}[htbp]
  \centering
  \caption{Lattice Parameter of tI50, oC120 and aP90 phases. Note that the experimental and theoretical melting point minimum corresponds to different pressures on the phase diagram. Thus the experimental lattice parameters were reported at nearly 120 GPa, and the predicted lattice parameters are calculated at 165 GPa. }\label{Table1}
  {\small
  \setlength{\tabcolsep}{3.5mm}
  \begin{adjustbox}{max width=\textwidth}
  \begin{tabular}{c|cc}
  \hline
  \hline
  Phase (Space Group)           &	Experimental Lattice Parameter\cite{NaPhaseDiagram2008Science} & Predicted Lattice Parameter\\
  \hline
  \multirow{3}{*}{$tI50$ ($I4/m$)}& \textbf{(117 GPa, 300 K)}      & \textbf{(165 GPa, 300 K)}    \\
                                  & a = b = 7.234(1) \AA           & a = b = 7.064(1) \AA         \\% delta: 0.023500138236107262
                                  & c= 9.193(3) \AA~(c/a=1.271)    & c = 8.973(1) \AA~(c/a=1.270) \\ %0.08% delta: 0.023931252039595223
\hline
  \multirow{4}{*}{$oC120$ ($Cmc2_1$)}& \textbf{(117 GPa, 300 K)}   & \textbf{(165 GPa, 300 K)}     \\
                                  & a = 6.65(1)  \AA               & a = 6.474(1) \AA              \\% delta: 0.026466165413533856
                                  & b = 13.17(1) \AA~(b/a = 1.980) & b = 12.928(2) \AA~(b/a=1.997) \\ % 0.86% delta: 0.018375094912680266
                                  & c = 13.12(2) \AA~(c/a = 1.973) & c = 12.857(2) \AA~(c/a=1.986) \\ % 0.66% delta: 0.020045731707317067
  \hline
  \multirow{6}{*}{$aP90$ ($P1$)}  & \textbf{(119 GPa, 300 K)}         &  \textbf{(165 GPa, 300 K)}     \\
                                  & a = 7.283(5)~(c/a = 2.216)       &  a = 7.138(1)  \AA~(c/a=2.214)  \\ % 0.09% delta: 0.01990937800357002
                                  & b = 7.366(5)~(c/b = 2.191)       &  b = 7.139(1)  \AA~(c/b=2.214)  \\ % 1.04% delta: 0.03081726853108871
                                  & c = 16.138(9)                     &  c = 15.805(2) \AA              \\% delta: 0.02063452720287532
                                  & $\alpha = \beta = 89.28(3)^\circ$ &  $\alpha = \beta = 89.90^\circ$ \\
                                  & $\gamma = 77.96(2)^\circ$         &  $\gamma=77.421(16)^\circ$      \\
  \hline
  \hline
  \end{tabular}
  \end{adjustbox}
  }
\end{table*}

\methodhead{Crystal structure searches and stability calculations}
Crystal structure searches were performed using CALYPSO\cite{calypso2010PRB,calypso2012CPC,Lvcalypso2012JCP,calypso2022JCP,WangHPCSD2026SciBull}, with DP-CSP replacing DFT for the local relaxation and energy evaluation of trial structures. The searches covered 20–60 and 88–92 atoms per primitive cell at 150 and 165 GPa. Candidate structures obtained from the searches were screened using DP-CSP and subsequently relaxed using DFT. Their relative enthalpies were calculated over 140–180 GPa and compared with those of established sodium phases. Details of the search and screening procedures are provided in the Supplemental Material.

% Avoid losing first-line width before the long PHONOPY name and citation.
\noindent Phonon dispersions were calculated using PHONOPY\cite{Phonopy2023JPCM} with DFT forces to assess the dynamical stability of selected candidates. For structures for which DFT phonon calculations were computationally prohibitive, structural persistence at finite temperature was examined using DP-FE molecular dynamics. In particular, simulations of \(aP90\) and \(oP92\) were performed at 165 GPa and 100 K for 500 ps, with atomic trajectories and mean-square displacements used to characterize atomic motion. Further details are provided in Supplemental Sec. III.

\methodhead{Gibbs free-energy calculations}
Gibbs free energies of sodium phases were calculated using DP-FE in two steps. First, Hamiltonian thermodynamic integration (HTI) established the free energy at a reference state \((T_0,P_0)\) by connecting each phase to a reference system: an Einstein crystal at 200 K for the solids and an ideal gas at 450 K for the liquid. Soft-core intermediate interactions were used to avoid close-contact singularities. Second, thermodynamic integration (TI) along isobars extended these reference free energies over 100–400 K for the solids and 300–450 K for the liquid, in 10 K intervals. Both steps were performed using the DPTI package with LAMMPS and a time step of 1 fs. Equilibration protocols, simulation lengths, reference-system parameters, and integration grids are detailed in Supplemental Sec. IV.

\methodhead{Melting-temperature calculations}
Melting temperatures were determined independently using solid–liquid coexistence simulations with DP-FE. For each phase, a structure relaxed at the target pressure was replicated to construct the simulation cell. A region at least 22 \AA~wide and containing at least 500 atoms was heated to 550 K to produce a liquid region while the remaining atoms were held fixed. The constraints were then released, and the entire system was simulated in the \(NPT\) ensemble at the target pressure and trial temperature. The melting temperature was identified from simulations in which solid and liquid regions persisted throughout the 350 ps trajectory. Details of the simulation cells and coexistence calculations are provided in Supplemental Sec. V.

%%%%%%% Insert Figures %%%%%%%%%
\begin{figure*}[t]
\centering
\includegraphics[width=\textwidth]{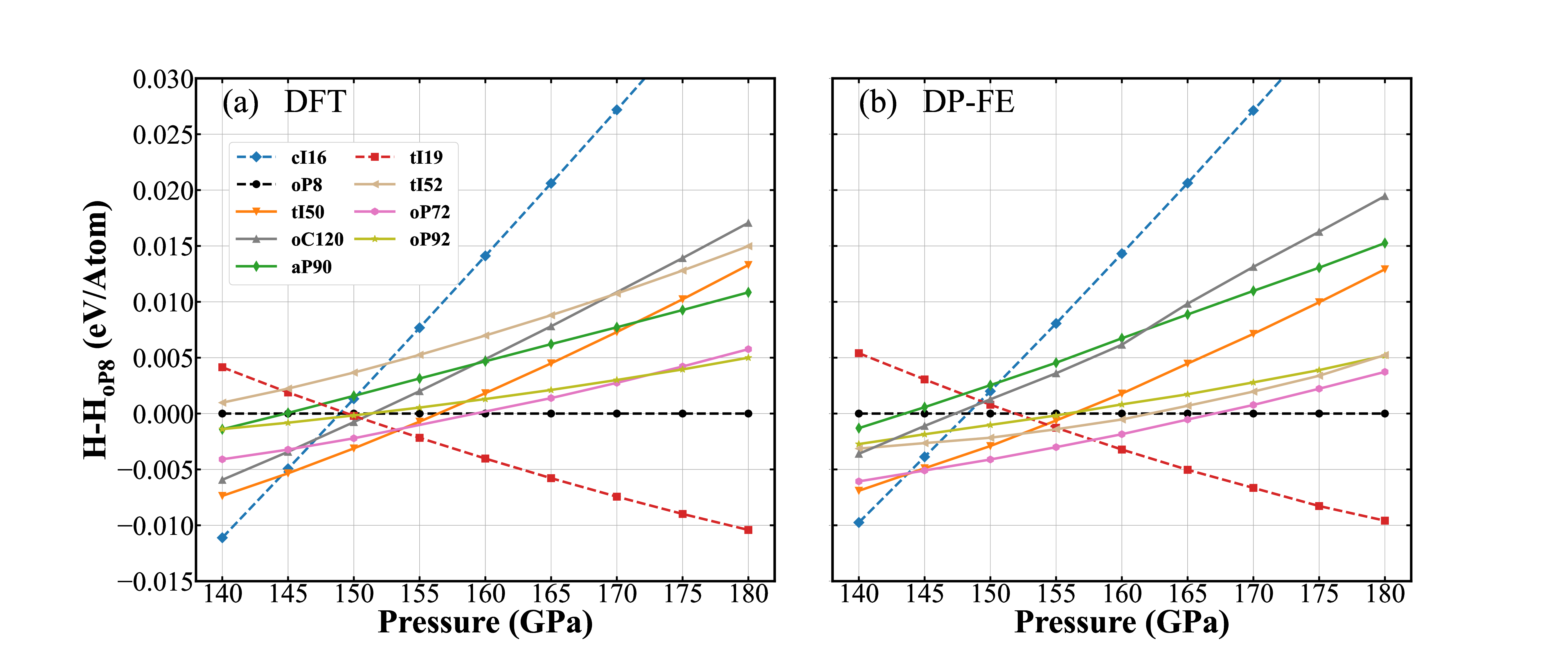}  % insert figure
\caption{Enthalpies of established and candidate sodium phases calculated with (a) DFT and (b) DP-FE.}
% TODO: verify the common energy-reference convention: oP8 is not at zero in panel (b).
\label{fig:Fig2}
\end{figure*}
%%%%%%%%%%%%%%%%%%%%%%%%%%%%%%%%%%

\section{Results and discussion}

To identify the complex crystal structures of sodium near its melting-curve minimum, we performed extensive crystal structure searches using CALYPSO coupled with the Deep Potential model DP-CSP~\cite{ConcurrentLearning2024PRB,WangLi2023NC}. The DP-CSP model enabled searches over more than 1.6 million configurations at 150 and 165 GPa, with 20--60 and 88--92 atoms per primitive cell. Candidate structures were screened by their DP-predicted enthalpies and subsequently refined using DFT. 
This procedure identified six low-enthalpy structures: candidate structural models for the experimentally observed $tI50$, $oC120$, and $aP90$ phases, and three additional competing structures, $tI52$, $oP72$, and $oP92$, shown in Fig.~\ref{fig:Fig1}.

\subsection{Crystal structures and lattice parameters}

The predicted $tI50$ structure is most easily visualized by focusing on the cage around a $2a$ atom. Four atoms form a square in the same plane as the central atom, while two further four-atom squares lie above and below it. The central square is rotated relative to the outer two, giving a cage of 12 surrounding atoms. These cages form part of a three-dimensional atomic framework, accompanied by columns of $8g$ atoms along $c$ with alternating short and long spacings. This atomic arrangement is consistent with the $I4/m$ model reported previously~\cite{YaoNa2025CommChem}.

The related $tI52$ structure retains the $I4/m$ symmetry and the three-square-ring, 12-coordinate cage motif, but the axial $2a$ orbit of $tI50$ is replaced by a $4e$ orbit. The $4e$ atom is not coplanar with the middle square, and the outer two rings lie at unequal distances above and below it. Thus, the difference between these two structures can be seen directly in the occupation of the fourfold axes and the shape of the occupied cages. Their common symmetry conceals a distinct local arrangement around the axial atoms. 

The $oC120$ candidate adopts $Cmc2_1$ symmetry and features both planar and corrugated atomic layers. Atoms at the $4a$ Wyckoff positions lie in planes at $x=0$ and $1/2$, whereas those at the $8b$ positions form intervening corrugated layers arranged symmetrically on either side of each $4a$ plane. And the $aP90$ candidate adopts $P1$ symmetry, with all 90 atoms occupying crystallographically independent sites, while retaining recognizable directional order. The atoms form undulating rows along $b$, with nonuniform displacements along $c$ and a staggered arrangement of adjacent rows.

The orthorhombic $oP72$ candidate shares the $Pnma$ space group with $oP8$ but exhibits a markedly different atomic arrangement. Only 12 of its 72 atoms lie in the corresponding $4c$ planes, while the remaining 60 occupy the intervening regions, including 56 atoms at general $8d$ positions that form corrugated layers. Its inversion center lies within the middle atomic layer. The $oP92$ candidate, with $Pbcm$ symmetry, admits a similar decomposition: 12 atoms lie in $4d$ planes at $z=1/4$ and $3/4$, while 80 atoms occupy general $8e$ positions between them. In this structure, the upper and lower halves are related by inversion about a center between the two central atomic layers. Full crystallographic parameters are provided in Supplemental Sec. VI.

To assess whether the predicted structures are compatible with the experimentally observed phases, we compare the lattice parameters of the $tI50$, $oC120$, and $aP90$ candidates with experimental data~\cite{NaPhaseDiagram2008Science} at 300 K and pressures near the theoretical melting-curve minima. As summarized in Table~\ref{Table1}, the calculated cell dimensions are systematically smaller by approximately 1.8--3.1\%, with the comparison involving different pressures near the theoretical and experimental melting minima. The axial proportions show closer agreement: for $tI50$, the calculated $c/a=1.270$ differs from the experimental value of 1.271 by only 0.08\%. For $oC120$, the calculated $b/a=1.997$ and $c/a=1.986$ reproduce the nearly twofold experimental ratios, with differences below 0.9\%. The $aP90$ candidate retains the elongated triclinic cell geometry, with $c/a$ and $c/b$ differing from experiment by 0.09\% and 1.04\%, respectively, and cell angles differing by no more than $0.70^\circ$. These agreements support compatibility with the indexed experimental cells.

\subsection{Enthalpy competition and dynamical stability}

We first assessed the thermodynamic competitiveness of the candidates by comparing their DFT enthalpies with those of the established $cI16$, $oP8$, and $tI19$ phases. The reference-phase enthalpies agree well with previous calculations~\cite{MaNa2009Nature}. As shown in Fig.~\ref{fig:Fig2}(a), several candidates compete near the calculated $cI16$--$oP8$ crossover. The $tI50$ structure lies below all three reference phases between approximately 145 and 153 GPa, while $oC120$ is favored over $cI16$ and $oP8$ within a narrower interval. Near the calculated $cI16$--$oP8$--$tI19$ enthalpy crossing, both $oC120$ and $aP90$ lie within 3 meV/atom of the lowest-enthalpy reference phase. The additional $oP72$ and $oP92$ structures are also competitive in this region, whereas $tI52$ follows a pressure dependence similar to that of $aP90$ but remains slightly higher in enthalpy.

The dynamical stability of the candidate structures was assessed using phonon calculations and molecular-dynamics simulations. For \(tI50\), \(tI52\), \(oP72\) and \(oC120\) phonon dispersions were calculated using DFT forces. % with \textcolor{red}{\(2\times2\times2\)} supercells. % TODO: verify the phonon supercell dimensions before reporting them.
No imaginary frequencies were observed at 165 GPa, supporting their dynamical stability within the harmonic approximation (Supplemental Fig. S2).

For the larger \(aP90\) and \(oP92\) structures, the high computational cost of DFT phonon calculations motivated an assessment of finite-temperature structural stability using DP-FE. The model closely reproduces their DFT enthalpy curves (Fig. 2b), with energy RMSEs of approximately 1.7–2.6 meV/atom for configurations sampled along their MD trajectories at 165 GPa and temperatures of 100 and 300 K (Supplemental Sec. II). Based on this agreement with DFT, we therefore used DP-FE to perform 500 ps DPMD simulations at 165 GPa and 100 K. 
Both structures retained their crystalline arrangements, with bounded mean-square displacements and atomic trajectories dominated by localized vibrations (Supplemental Fig. S3), indicating structural stability over the simulated time scale.

With several candidates close in enthalpy and retaining their structures under the conditions examined, an important remaining question is how temperature affects their relative stability. We address this question through Gibbs free-energy calculations.

\begin{figure*}[t]
\centering
\includegraphics[width=1.0\textwidth]{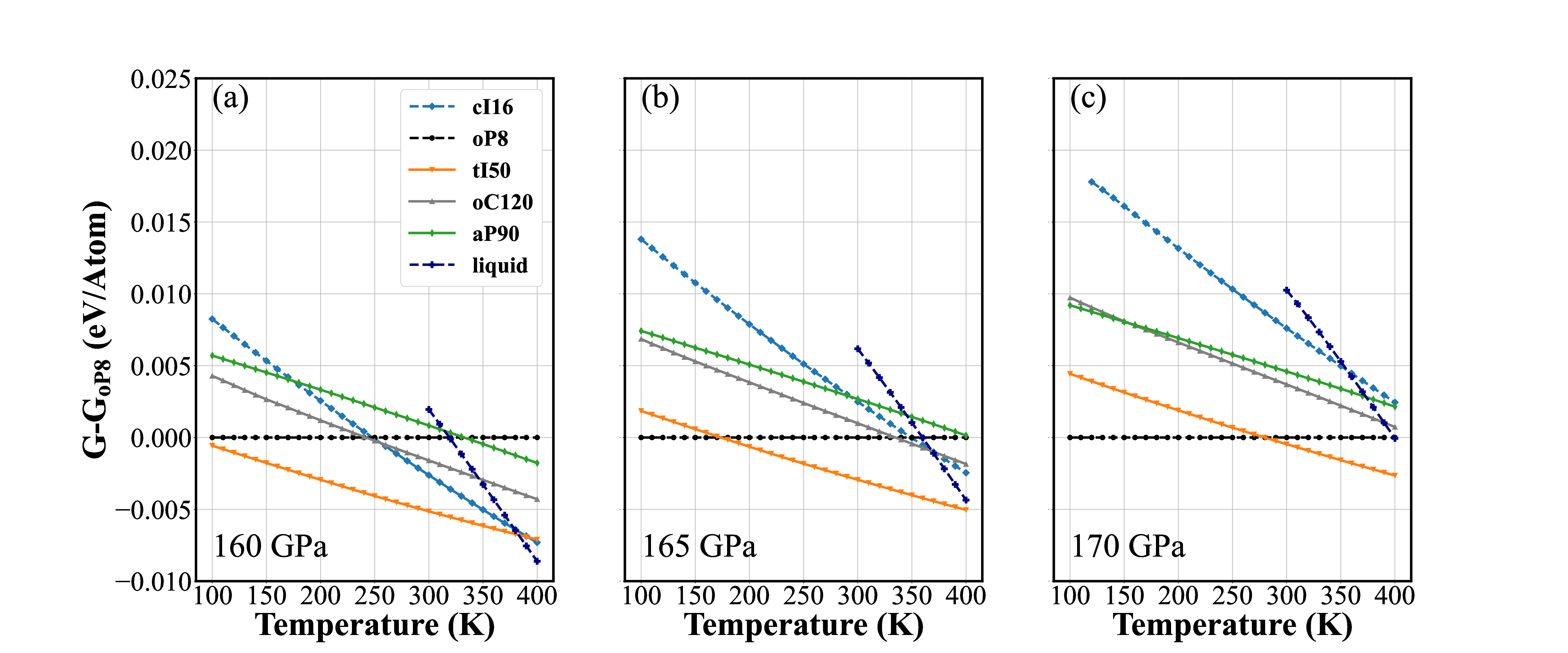}
\caption{Gibbs free energies calculated with DP-FE relative to $oP8$ at (a) 160, (b) 165, and (c) 170 GPa. Solid lines denote the established $cI16$ and $oP8$ phases and the liquid; dashed lines with symbols denote the $tI50$, $oC120$, and $aP90$ candidates. A decrease with temperature indicates an entropy advantage relative to $oP8$, while equilibrium stability requires comparison with every competing phase.}
\label{fig:Fig3}
\end{figure*}

\subsection{Finite-temperature thermodynamics stability}

We further examined the finite-temperature thermodynamic behavior of the candidate structures by calculating their Gibbs free energies near the melting point minimum through thermodynamic integration using DP-FE. Fig.~\ref{fig:Fig3} compares the free energies of \(tI50\), \(oC120\), \(aP90\), \(cI16\), \(oP8\), and the liquid at 160, 165, and 170 GPa, with \(oP8\) taken as the reference. The relative free energies of \(tI50\) and \(oC120\) decrease with increasing temperature at all three pressures, although their ordering relative to \(cI16\) varies. At 160 GPa, \(tI50\) is favored at lower temperatures, while \(cI16\) falls below it upon heating. At 165 and 170 GPa, \(tI50\) remains below \(cI16\) and \(oP8\) as melting is approached. The \(oC120\) candidate is closely competitive with the established phases, particularly at 165 GPa, where its Gibbs free energy becomes slightly lower than those of both \(cI16\) and \(oP8\) over a narrow temperature interval before melting. The \(aP90\) candidate also remains within a few meV/atom of the reference phases near the calculated melting minimum, although it does not become the lowest-free-energy structure in the comparisons shown.

The free-energy differences among the competing phases are only a few meV/atom, making their relative ordering sensitive to model accuracy. Validation against DFT yields phase-resolved energy RMSEs of approximately 1.1–2.6 meV/atom for the five solid phases compared here at 165 GPa and 300 K (Supplemental Table S1). 
% \textcolor{red}{And the error caused by statistical variations in DPMD and numerical integrations is no higher than 0.6 meV/atom}.
Although these energy RMSEs do not directly quantify the uncertainty in Gibbs free energies, the phase-dependent energy biases observed in the validation may affect the ordering of closely competing phases. In particular, DP-FE underestimates the energy of \(cI16\) more strongly than that of \(oC120\), as shown in Fig.~S1, which may preferentially stabilize \(cI16\) relative to \(oC120\) in the model predictions. Given these phase-dependent model biases, the small free-energy penalties found in the calculations do not exclude \(oC120\) and \(aP90\) from thermodynamic competition with the reference phases near the calculated melting minimum.

\begin{figure*}[t]
\centering
\includegraphics[width=0.7\textwidth]{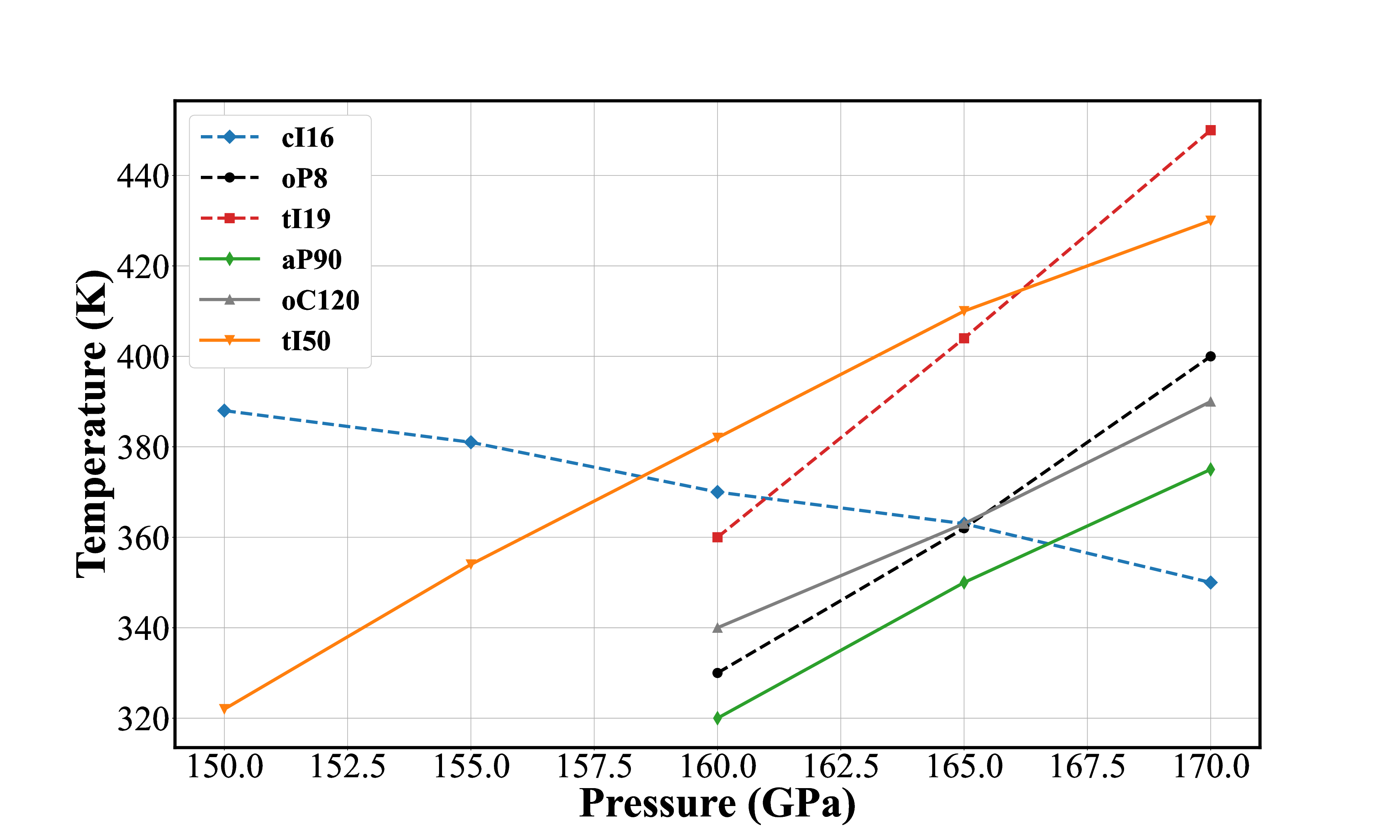}
\caption{Melting curves obtained from solid--liquid coexistence simulations with DP-FE. Dashed lines denote the established $cI16$, $oP8$, and $tI19$ phases; solid lines denote the $tI50$, $oC120$, and $aP90$ candidates. The $cI16$--$oP8$ curves define a reference minimum near 165 GPa. Including $tI50$ raises the upper envelope of the compared curves in this region and shifts its minimum to approximately 158 GPa.}
\label{fig:Fig4}
\end{figure*}

\subsection{Melting competition and implications for the phase diagram}

The $tI19$ phase exhibits chain melting during DPMD simulations, with rapid guest-atom diffusion along channels parallel to $c$. Similar behavior occurs in $tI52$. This motion distinguishes these structures from crystals in which all atoms remain localized about fixed lattice sites, and could introduce an additional entropic contribution. The HTI protocol used here does not accommodate this diffusion, so its contribution and the Gibbs free energy of $tI19$ and $tI52$ were not evaluated. Solid--liquid coexistence simulations provide a complementary comparison that allows $tI19$ and $tI52$ to be retained. The resulting melting curves are shown in Fig.~\ref{fig:Fig4}, with details in Supplemental Sec. V.

The $cI16$ and $oP8$ curves define a reference melting minimum near 165 GPa and 360 K, consistent with the free-energy calculations. At this pressure, $tI50$ melts at approximately 410 K and remains competitive with $tI19$; on further compression, the $tI19$ melting curve rises above that of $tI50$. In contrast, $oC120$ melts at approximately 362 K, indistinguishable from the reference minimum within the estimated coexistence uncertainty of about 5 K, and $aP90$ melts at approximately 350 K. Their proximity to the $cI16$--$oP8$ reference minimum supports close competition with those reference phases.

The additional structures extend this competition (Supplemental Fig.~S4). At 165 GPa, $oP72$ has a melting temperature of approximately 418 K, slightly above $tI50$, while $tI52$ and $oP92$ melt near 390--395 K, above the $cI16$--$oP8$ reference but below $tI50$. At 170 GPa, $oP72$ and $tI19$ approach 450 K, exceeding the approximately 430 K melting temperature of $tI50$. These comparisons show that the three additional low-enthalpy structures can also compete near melting, but do not establish equilibrium stability fields. In particular, DP-FE underestimates the energy of $oP72$ by approximately 4 meV/atom at 300 K (Supplemental Fig.~S1), which can artificially favor the solid and raise its predicted melting temperature.

However, a substantial quantitative discrepancy with experiment remains: the reference melting minimum defined by $cI16$ and $oP8$ lies near 165 GPa, well above the experimental minimum near 118 GPa. The DFT enthalpy crossing between these phases also occurs at an elevated pressure of approximately 149 GPa, suggesting that the discrepancy does not arise solely from the DP description of finite-temperature behavior. Similar discrepancies between calculated and experimentally observed transition pressures have been reported previously~\cite{MaNa2009Nature,YaoNa2025CommChem}.

For example, Ref.~\cite{MaNa2009Nature} predicts an athermal $tI19 \rightarrow hP4$ transition in sodium at 260 GPa, approximately 60 GPa above the experimental transition pressure, and attributes this discrepancy to the tendency of DFT to overstabilize metallic states. However, both $cI16$ and $oP8$ are metallic in the pressure range considered here, so preferential stabilization of metallic over insulating states does not directly explain the present discrepancy.

Nevertheless, exchange-correlation functional dependence remains a plausible contributor. A relevant precedent is dense lithium, where PBE calculations place $oC88$ above $cI16$ in free energy within the experimentally observed stability range of $oC88$~\cite{WangLi2023NC}, even after nuclear quantum and anharmonic effects are included~\cite{LioC88cI16HSE2025PRL}. An HSE06 single-point correction at 70 GPa lowers the relative energy of $oC88$ by approximately 6.17 meV/atom, bringing its predicted stability into closer agreement with experiment~\cite{LioC88cI16HSE2025PRL}. This example illustrates that functional choice can alter the relative stability of closely competing alkali-metal phases. Calculations using alternative exchange-correlation functionals may therefore help assess, and potentially reduce, the pressure offset found here. Whether such corrections shift the sodium phase boundaries toward experiment, and by how much, remains to be established.

\section{Conclusion}

In summary, we combine CALYPSO crystal structure prediction with Deep Potential models developed through concurrent learning to explore the complex crystalline structures of sodium near its melting minimum. Searches over more than 1.6 million configurations yield candidate atomic arrangements for the experimentally unresolved $oC120$ and $aP90$ phases, recover the previously proposed $tI50$ structure, and identify three additional structures, $tI52$, $oP72$, and $oP92$. The agreement of the predicted lattice geometries with experimental constraints, together with the finite-temperature assessments, supports the proposed models as plausible candidates for the complex phases observed in this region. These findings extend the known structural possibilities of dense sodium and provide concrete atomic models for interpreting the experimental observations.

The pressure offset between the calculated and experimental melting minima remains unresolved. The functional dependence of phase stability demonstrated in dense lithium suggests that the exchange-correlation approximation may contribute to this discrepancy, a possibility that warrants further investigation in sodium. Extending the present approach to larger unit cells, including those relevant to the unresolved $mP512$ phase, together with improved electronic-structure calculations, may help complete the structural picture of sodium near its melting minimum.

\section*{acknowledgments}
This work was supported by the Fundamental and Interdisciplinary Disciplines Breakthrough Plan of the Ministry of Education of China (Grant No. JYB2025XDXM403), Scientific and Technological Innovation Project of Changbaishan Laboratory, Jilin Province (Grant No. CBS2025007-01), and the National Natural Science Foundation of China (Grant Nos. 12374005, T2225013, 52288102 and U22A2098). Zhenyu Wang. was supported by the Postdoctoral Fellowship Program of CPSF (Grant No. GZC20252243). Part of the calculation was performed in the high-performance computing center of Jilin University.

\section*{Data Availability}
CALYPSO code is free for academic use, by registering at \url{http://www.calypso.cn}. LAMMPS, DP-GEN, DeepMD-kit and DPTI are free and open-source codes available at \url{https://lammps.sandia.gov} and \url{https://deepmodeling.com}, respectively. The other scripts are available from the authors upon request.

\nocite{*}

\bibliographystyle{apsrev4-2}
\bibliography{ref}

\end{document}